\documentstyle[12pt]{article}

\input epsf.sty
\newdimen\psfigsize
\def\psfigure#1 #2 #3 #4 #5{
    \begin{figure}[tbp]
    \vbox{
    \null\hskip#2\epsfxsize=#1 \epsfbox{#4}
    \vskip 10truept
    \caption {#5 \label{#3}}
    \vskip 0.1truein plus0.2truein}
    \end{figure}
}
\def\pspagefigure#1 #2 #3 #4 #5{
    \begin{figure}[p]
    \vbox{
    \null\hskip#2\epsfxsize=#1 \epsfbox[0 0 4096 4096]{#4}
    \vskip 10truept
    \caption {#5 \label{#3}}
    \vskip 0.1truein plus0.2truein}
    \end{figure}
}
\def\psoddfigure#1 #2 #3 #4 #5 #6{
    \begin{figure}[tbhp]
    \vbox{
    \null\hskip#3\epsfxsize=#1 \epsfbox[0 0 4096 4096]{#5}
    \vskip -#1 \vskip #2 \vskip 10truept
    \vskip 10truept
    \caption {#6 \label{#4}}
    \vskip 0.1truein plus0.2truein}
    \end{figure}
}
\def\figurespace#1 #2 #3 #4 {
    \begin{figure}[tbhp]
    \vbox{
    \psfigsize=#1truein
    \vskip \psfigsize
    \vskip 10truept
    \caption {#4 \label{#3}}
    \vskip 0.1truein plus0.2truein}
    \end{figure}
}
\def\gnufigure#1 #2 #3 #4 #5 #6{
    \begin{figure}[tbhp]
    \vbox{
    \null\hskip#3\epsfxsize=#1 \epsfbox{#5}
    \vskip -#1 \vskip #2 \vskip 10truept
    \vskip 10truept
    \hbox{\null\hskip 1.0in \parbox[t]{4.5in}{ \caption {#6 \label{#4}} } }
    \vskip 0.1truein plus0.2truein}
    \end{figure}
}

\def\etal{{\it et al.}}

\def\Dslash{\mathop{\not\!\! D}}

\def\LP{\left(}		
\def\RP{\right)}	
\def\LB{\left\{}	
\def\RB{\right\}}	
\def\PAR#1#2{ {{\partial #1}\over{\partial #2}} }

\def\BE{\begin{equation}}
\def\EE{\end{equation}}
\def\BEA{\begin{eqnarray}}
\def\EEA{\end{eqnarray}}
\def\EL{\nonumber\\}

\newcommand{\la}[1]{\label{#1}}

\begin{document}

\begin{titlepage}
\baselineskip=16pt
\rightline{\bf hep-lat/9805009}
\baselineskip=20pt plus 1pt
\vskip 1.5cm

\centerline{\Large \bf Testing improved actions for dynamical Kogut-Susskind quarks}
\bigskip
\centerline{\bf Kostas~Orginos }
\centerline{\it
Department of Physics, University of Arizona, Tucson, AZ 85721, USA}
\bigskip
\centerline{\bf Doug~Toussaint }
\centerline{\it
Department of Physics, University of Arizona, Tucson, AZ 85721, USA}
\centerline{\it and}
\centerline{\it
Center for Computational Physics, University of Tsukuba, Ibaraki 305, Japan }
\vskip 0.5in
\centerline{ The MILC Collaboration }

\narrower
We extend tests of ``Naik'' and ``fat link'' improvements of the
Kogut-Susskind quark action to full QCD simulations, and verify that the
improvements previously demonstrated in the quenched approximation apply
also to dynamical quark simulations.  We extend the study of flavor
symmetry improvement to the complete set of pions, and find that the
nonlocal pions are significantly heavier than the local non-Goldstone
pion.  These results can be used to estimate the lattice spacing
necessary for realistic simulations with this action.
\end{titlepage}

\section{Introduction}

Because of the much larger computer time required to generate sample
lattices with small lattice spacing in full QCD,
improved actions are even more important to full QCD simulations than
to quenched simulations.  However, because of this same high cost of
computing, the effects of improving the action have mostly been tested
in the quenched approximation.

For Wilson quarks, a Symanzik improved gauge action with a clover quark
action\cite{MILC_WILSON_THERMO,BIELEFELD_WILSON_THERMO}
and a renormalization group motivated improved gauge action with
the conventional Wilson quark action\cite{JLQCD_WILSON_THERMO}
have been used in studies of full QCD thermodynamics.
These studies showed significant changes in the
physics relative to simulations with the conventional action.

Studies of improvements with Kogut-Susskind quarks have been slower
to start, possibly because it is less urgent to cure the $o(a^2)$ errors
in this formulation than the $o(a)$ errors in the conventional Wilson
quark action.  Kogut-Susskind quarks are preferred for studies of
light quark thermodynamics because of their remnant exact chiral
symmetry, but studies of quantities such as the equation of state have
been severely limited by lattice effects\cite{NT6_EOS}.

Two improvements to the Kogut-Susskind quark action have been
investigated.  Adding a third nearest neighbor term\cite{NAIK}
improves the quark dispersion relation and ``free field'' quantities
such as the Stefan-Boltzmann law on the lattice.  The effects of this
improvement on the quenched hadron spectrum have been studied by the
MILC collaboration\cite{MILC_NAIK}, where it was found that it
improves rotational symmetry as measured by the dispersion relation of
the pions, but has only small effects on flavor symmetry breaking or
spin splittings of the hadrons.  Dynamical high temperature studies
using the Naik action with four flavors of quarks have been done by
the Bielefeld group\cite{BIELEFELD_NAIK}.  A second kind of
improvement --- averaging over paths in the parallel transport of the
quark fields, or ``fattening'', does improve the flavor
symmetry\cite{MILC_FATLINKS}.  This can be understood as a suppression
of the coupling to high momentum gluons which scatter quarks from one
corner of the Brillouin zone to another\cite{LEPAGE_TSUKUBA}, and this
understanding has been used as a guide to further reducing flavor
symmetry breaking by more extensive fattening\cite{ILLINOIS_FAT}.
Recently, in\cite{TD_AH_TK} it was shown that in SU(2) quenched
spectroscopy the local non-Goldstone pion becomes almost degenerate
with the Goldstone pion after smoothing the gauge fields with RG ``cycling''.

This work explores the effects of these improvements of the
Kogut-Susskind quark action, together with a Symanzik improved gauge
action, on simulations with dynamical quarks.  One motivation for this
is to make sure that the promising results from the quenched
approximation also apply to full QCD, and a second motivation is to
begin developing a base of experience that can be used to estimate the
quality of a physics simulation done with these actions at a particular
lattice spacing, or alternatively to estimate the lattice spacing
required for a simulation of some particular quality.
We study improvement in rotational invariance diagnosed through the
pion dispersion relation, and flavor symmetry restoration from the zero
momentum masses of the pions.  In addition to the local non-Goldstone
pion, the ``$\pi_2$'', we look also at the nonlocal pions
\cite{GOLTERMAN_MESONS,ALLPIONMASSES}.
No attempt is made at this stage to study whether these improved actions
show better scaling of mass ratios.

\section{The simulations}

The gauge action used in most of these simulations was a Symanzik improved
gauge action\cite{Symanzik83,LuscherWeisz,LepageMackenzie,Alford95}.  This is
the same gauge action used in the MILC studies of improvement for the
quenched spectrum\cite{MILC_NAIK}.
\BE
  S_G = \LP \frac{\beta_{imp}}{3} \RP
   \LP  \sum_{x;\mu<\nu} (P_{\mu\nu})
        - \frac{1}{20\,u_0^2}(1+0.4805\,\alpha_s)
        \sum_{x;\mu\neq\nu} (R_{\mu\nu})
        - \frac{1}{u_0^2} 0.03325 \,\alpha_s
        \sum_{x;\mu<\nu<\sigma} ( C_{\mu\nu\sigma})
   \RP
 \la{gaugeaction}
\EE
where $P$ is the standard plaquette in the $\mu,\nu$ -plane, and $R$
and $C$ denote the real part of the trace of the ordered product of
SU(3) link matrices along $1\times 2$ rectangles and $1\times 1 \times
1$ paths, respectively. Here $\beta_{imp} = 10/g^2$,
and $\alpha_s= -4\log(u_0)/3.0684$.
Because these are exploratory calculations, no effort was made to tune
$u_0$.  Instead, the value found in Ref.~\cite{MILC_NAIK} at $\beta_{imp}=7.1$,
$u_0=0.8441$ was used.

The conventional one plaquette action was used in some simulations
for comparison.
\BE
  S_G = \LP \frac{\beta_{conv}}{3} \RP \LP  \sum_{x;\mu<\nu} (P_{\mu\nu}) \RP
\la{convgaugeaction}
\EE
Here $\beta_{conv} = 6/g^2$.

The quark action used the hopping matrix $2m + \Dslash$ with $\Dslash$
parameterized as

\BEA
\Dslash(x,y) &=& \sum_\mu \eta_\mu(x) \EL
\bigg(
&& c_1  \LP U_\mu(x)\delta_{y,x+\hat\mu} -
U^\dagger_\mu(x-\hat\mu)\delta_{y,x-\hat\mu} \RP \EL
&+& c_3 \LP U_\mu(x)U_\mu(x+\hat\mu)U_\mu(x+2\hat\mu) \delta_{y,x+3\hat\mu}
- {\rm ``backwards\ Naik\ term"} \RP \EL
&+& w_3 \sum_{\nu\neq\mu} \Big( 
\LB U_\nu(x)U_\mu(x+\hat\nu)U^\dagger_\nu(x+\hat\mu) +
    U^\dagger_\nu(x-\hat\nu)U_\mu(x-\hat\nu)U_\nu(x-\hat\nu+\hat\mu) 
\RB \delta_{y,x+\hat\mu} \EL
 &-& {\rm ``backwards\ staple\ term"}
\Big)
\bigg)
\EEA
Here $c_1$ is the coefficient of the conventional single-link term,
$c_3$ is the coefficient of the third nearest neighbor (Naik) term,
and $w_3$ is the coefficient of the ``fat link'', or staple term.
Note that this parameterization of the fattening is designed to be
convenient for full QCD updating, and differs from the parameterization
used in Ref.~\cite{MILC_FATLINKS}.

Four different valence quark actions were used in computing the
spectrum.
The conventional ``one-link'' quark action has $c_1=1$ and $c_3=w_3=0$.
The ``Naik'' action has $c_1=9/8$, $c_3=-1/24$ and $w_3=0$.
For the ``fat-link'' action we arbitrarily weight each staple with
one half the weight of the single link path, and normalize so that the
total weight of the one link path is one: $c_1 = 1/4$, $c_3=0$, and
$w_3=1/8$.  In Ref.~\cite{MILC_FATLINKS} it was found that the
improvement in flavor symmetry is fairly insensitive to the exact weight of
the staples.  Finally, for the ``one+fat+Naik'' action, we keep the
relative weight of the staple to the one link term at one half, but
normalize the total of the single link plus six staples to $9/8$:
$c_1 = (9/8)(1/4)$, $c_3=-1/24$ and $w_3=(9/8)(1/8)$.  Notice that we
have not tadpole improved the quark action by including a factor of
$1/u_0^2$ in the third nearest neighbor coupling, as was done in
Ref.~\cite{MILC_NAIK}.  This would have been reasonable for the Naik
action, but for the one+fat+Naik it would have made much less
difference, since the bulk of the coupling to the nearest neighbor sites
also comes from three link paths, which presumably are boosted by the
same factor.  Here the main effect would have been a renormalization of
the mass.

The gauge action and gauge force computations used a generalized gauge
action code written by Tom DeGrand and Anna Hasenfratz\cite{GAUGE_FORCE_CODE}.
For the inversions
of the fermion matrix, the conjugate gradient algorithm was used.  The
products of links connecting to the third nearest neighbor sites and
the sum of the single link plus staples were precomputed before
starting the conjugate gradient, so the computer time per iteration was
about twice that of the conventional one-link quark action.  The fermion
force can be computed by a straightforward generalization of the
algorithm described in \cite{R_ALGORITHM}.  The actions used here are
all in the class of actions where the fermion matrix is $2m$ times the
identity matrix plus a $\Dslash$ operator which couples even sites to
odd sites.  For these actions the Hermitian positive definite operator
$M^\dagger M$ is block diagonal, coupling even sites only to even sites.
Thus, in the standard way, the pseudofermion action can be written
without further doubling the number of flavors from four to eight as
\BE S_F = \phi_e^\dagger \frac{1}{M^\dagger M} \phi_e \ \ \ ,\EE
where $\phi_e$ is a pseudofermion defined on the even lattice sites
($x+y+z+t$ is even).
(For simplicity, we sketch the force computation for four flavors of
quarks with the hybrid Monte Carlo algorithm; adapting it to two flavors
is done in the standard way\cite{R_ALGORITHM}.
Roughly,
\BEA F_f
&\approx& \PAR{}{U} \langle \phi_e | \frac{1}{M^\dagger M} | \phi_e \rangle \\
&=& \langle\phi_e|\frac{1}{M^\dagger M} \PAR{}{U}(M^\dagger M)
\frac{1}{M^\dagger M} | \phi_e \rangle \\
&=& \langle X_e |  \PAR{}{U}(M^\dagger M) | X_e \rangle
\la{fforceeq}
\EEA
where $\phi_e$ is restricted to even lattice sites and $X_e$ is
$(M^\dagger M)^{-1}\phi_e$, also restricted to even sites.  The diagonal
(mass) term in $M^\dagger M$ doesn't contribute to the force, so by
defining $X_o$ on odd lattice sites to be $\Dslash X_e$, this can
be written as
\BE \langle X_o | \PAR{M}{U} | X_e \rangle + \langle X_e |
\PAR{M^\dagger}{U} | X_o \rangle \ \ \ .\EE
Except for a minus sign from the adjoint, the two terms can
be treated identically.
Now $M$ depends on a particular link matrix $U_\mu(x)$ through every
path in the quark action which contains this link.  After projecting out
the traceless anti-Hermitian part, the terms where the link is going
backwards ($U^\dagger$ appears in $M$) just give a factor of two.
So, with $\Dslash X_e$ stored in the odd sites of $X$, the force
computation involves
\begin{itemize}
\item{} Compute $X_e = (M^\dagger M)^{-1} \phi_e$ on even sites.
\item{} Compute $X_o = \Dslash X_e$ on odd sites
\item{} For each path in the quark action, for each link going in
direction $+ \hat\mu$, calculate a contribution to the ``force'', or
change in the momentum associated with the link:
\begin{itemize}
\item{} Parallel transport $X$ along the path from each end
of the path to the site of the momentum
being updated.  This involves multiplying by $U$ or $U^\dagger$, and by
the Kogut-Susskind phase factors $\eta_\nu(x)$.
\item{}Take the traceless anti-Hermitian part of $|F\rangle\langle B|$,
where $F$ and $B$ are $X$ transported from the forwards and backwards
ends of the path.
\item{}Multiply by $-1$ if the link is an odd numbered link in the path,
and by $-1$ if you are updating the momentum on an odd site.  This takes
care of the minus signs coming from the adjoint in the last line of
Eq.~\ref{fforceeq}.
\item{}Since the $\eta_\mu$ contribute an unwanted minus sign for every
plaquette enclosed by a loop, multiply by $-1$ for each plaquette
enclosed by a loop consisting of the path being considered and some
standard path.  For example, when the Kogut-Susskind phases are absorbed
into the link matrices, the staple contribution to the fat link gets a
minus sign relative to the single link connection.
\item{}Multiply by the coefficient of the path, which includes a minus
sign for paths that are part of the ``backwards'' part of $\Dslash$.

\end{itemize}
\end{itemize}

Some optimizations are possible, since several different paths passing
through a site may share the same forwards or backwards parts and
vectors transported to one site and used in the force computation there
can be subsequently transported to the next site along the path.  
The timings of the various parts of the computation depend on many
factors such as the machine used, the lattice sizes, and how carefully
each part is optimized.  As an example, on an $8^4$ lattice on an
alpha workstation with this improved action
the gauge force computation requires 24.6 seconds, the
fermion force computation 5.4 seconds and the conjugate gradient 0.183
seconds per iteration.  About 400 conjugate gradient iterations were
required at $\beta_{imp}=7.5$ and $am_q=0.015$.

Since getting good acceptance in the hybrid Monte Carlo algorithm is
a good test of the program, 
some four flavor results on small lattices were done first.
While these were intended just to be test runs, because they had large
lattice spacing and larger quark mass than the two flavor runs they
ended up showing the largest effects on the pion dispersion relation, so
some of the four flavor results for the pion dispersion
relation are reported. Two
flavor runs with the improved action were done at $\beta_{imp}=7.1$,
$7.3$ and $7.5$.  Some of the conventional action results needed for
comparison can be found in the literature, but some conventional action lattices
were generated at $\beta_{conv}=5.5$.  This was done to get a reasonable
match of lattice spacing and lattice size to the $\beta_{imp}=7.3$
improved action runs and to get the masses of all of the pions.
For all of the sample sets that were generated,
the meson spectrum was calculated with both the conventional and
improved valence actions.  In some cases valence actions where the Naik
and the fattening were turned on separately were also used.
Several local meson propagators were calculated at nonzero momenta, but
the Goldstone pion gave by far the most accurate results, and our
analysis of rotational symmetry will concentrate on these propagators.

\section{Results}

The largest data sets are two flavor runs at $\beta_{imp}=7.1$, $7.3$
and $7.5$, together with conventional action runs at $\beta_{conv}=5.5$
chosen to match the $\beta_{imp}=7.3$ runs.  At $\beta_{imp}=7.1$ a
single mass value was run, at $m_\pi/m_\rho=0.51$ for the conventional action
and  $m_\pi/m_\rho=0.57$ for the fat+Naik action.  At the two larger
$\beta_{imp}$ values, two quark masses were used which bracket
$m_\pi/m_\rho=0.55$, so it is possible to interpolate to this
arbitrary value for comparing the different actions.
Also, to compare the lattice spacings of the different runs we can
use the $\rho_2$ mass interpolated to $m_\pi/m_\rho=0.55$ as an
arbitrary length standard, and one which is more accessible than an
extrapolation to the physical light quark mass.
(The $\rho_2$ is used instead of the $\rho$ since at large
lattice spacings it often gives a better signal.)

Table~I contains the pion masses, including both local and nonlocal
pions at zero momentum, and the Goldstone pion at
momenta $\frac{2\pi}{L}(0,0,0)$,
$\frac{2\pi}{L}(0,0,1)$ and $\frac{2\pi}{L}(0,1,1)$.
The $\rho_2$ mass is also tabulated, as it will be used to set the
scale.

The absence of rotational symmetry on the lattice leads to meson dispersion
relations that differ from the $E(\vec k)^2 = m^2+\vec k^2$ expected in the
continuum.  The dimensionless parameter ``speed of light squared'', which
should be one, is used to parameterize the deviations from the correct
dispersion relation,
\BE c^2 = \frac{E^2(\vec k) - E^2(\vec 0)}{\vec k^2} \ \ \ .\EE
Table~II shows $c^2$ for the Goldstone pion at the
two smallest lattice momenta in selected runs.
While the two flavor runs with smaller quark masses
are relevant for estimating the lattice
spacing required for physical simulations, improvement of the dispersion
relation is naturally most dramatic in the runs where it is worst with
the unimproved action, namely at large lattice spacing and large quark
mass.  Therefore Table~II also contains a four flavor
run at quark mass $am_q=0.1$, corresponding to $m_\pi/m_\rho=0.76$ or
$0.82$ with the conventional and improved actions respectively.
Ideally, all of these numbers would be quoted at the same physical
momentum.  A crude approximation to this was achieved by increasing
the spatial lattice size in units of $a$ as $\beta$ was increased,
hoping to keep the physical lattice size constant.  The extent to which
this was achieved is shown by expressing the unit of lattice momentum,
$2\pi/L$ in units of the $\rho_2$ mass interpolated to
$m_\pi/m_\rho=0.55$, and is included in this table.
Since $c^2$ involves the difference of the pion energy at zero and nonzero momentum, a jackknife
analysis was done to estimate the statistical error on $c^2$ at
$\beta=7.3$ and $7.5$.
Errors estimated from the jackknife analysis were not significantly
different than errors estimated naively from the statistical error in
the nonzero momentum pion energies.
The improved action $10/g^2=7.3$ and conventional action $6/g^2=5.5$
results in Tables I and II contain the non-Goldstone pion masses
and speed of light for the Goldstone pion with the
fat-link and Naik improvements turned on separately.  Within the fairly
poor statistics, this is consistent with the quenched result of
Ref.~\cite{MILC_NAIK} that, as expected, it is the Naik term that is
responsible for the improvement in the dispersion relation.


\newpage
\begin{center}\begin{tabular}{llrlllllll}

$\beta$	& $m_q$	& $N_s$	& $S_V$	& $\rho_2$ & $\pi_0$ & $\pi_1$ & $\pi_2$ & $\pi_3$ & $\vec p_\pi$ \\
& & & & & $\gamma_5\otimes\gamma_5       $           & $\gamma_5\otimes\gamma_i\gamma_5$        &
	$\gamma_5\otimes\gamma_i\gamma_0$     & $\gamma_5\otimes\gamma_0$         & $(0,0,1)$ \\
& & & & & $\gamma_0\gamma_5\otimes\gamma_0\gamma_5$ & $\gamma_0\gamma_5\otimes\gamma_i\gamma_j$ &
	$\gamma_0\gamma_5\otimes\gamma_i$    & $\gamma_0\gamma_5\otimes{\bf 1}$   & $(0,1,1)$ \\
\hline
$7.1_I$ & 0.05	& 10	& OL  & 1.180(26) & 0.6033(4) & 1.035(8) & 1.170(12)& 1.16(7) & 0.831(3) \\
        &     	&  	&     &           & 1.013(10) & 1.127(25)& 1.25(2)  & 1.21(3) & 0.980(4) \\
	&	&	& OFN & 1.080(14) & 0.6180(7) & 0.772(2) & 0.842(6) & 0.911(8)& 0.865(5) \\
	&	&	&     &           & 0.765(3)  & 0.846(6) & 0.920(10)& 0.96(2) & 1.067(14) \\
\hline
$7.3_I$ & 0.020 & 12	& OL  & 0.861(15) & 0.3985(7) & 0.635(6) & 0.719(13)& 0.74(2) & 0.642(4) \\
        &       &   	&     &           & 0.636(11) & 0.69(1)  & 0.73(3)  & 0.74(3) & 0.800(6) \\
	&	&	& ON  & 0.81(2) & 0.383(1) & 0.620(4) & --- & --- & 0.643(4) \\
	&	&	&     &         & 0.60(1)  & 0.67(1)  & --- & --- & 0.81(1) \\
	&	&	& OF  & 0.73(2) & 0.373(1) & 0.455(3) & --- & --- & 0.629(8) \\
	&	&	&     &         & 0.460(5) & 0.496(5) & --- & --- & 0.85(2) \\
	&	&	& OFN & 0.722(11) & 0.360(1) & 0.443(4) & 0.490(4) & 0.532(5) &  0.63(1) \\
	&	&	&     &           & 0.445(4) & 0.488(6) & 0.541(7) & 0.572(9) & 0.86(2) \\
\hline
$5.5_C$ & 0.015 & 12	& OL  & 0.71(1) & 0.3513(9) & 0.536(5)  & 0.584(9) & 0.594(13)& 0.619(7) \\
        &       &   	&     &           & 0.541(8) & 0.581(10)& 0.586(8) & 0.594(5) & 0.79(1) \\
	&	&	& ON  & 0.71(1) & 0.341(1) & 0.517(4) & --- & --- & 0.619(9) \\
	&	&	&     &         & 0.53(1)  & 0.570(10) & --- & --- & 0.81(1) \\
	&	&	& OF  & 0.61(1) & 0.317(2) & 0.369(4) & --- & --- & 0.585(12)\\
	&	&	&     &         & 0.390(6) & 0.401(7) & --- & --- & 0.84(2)  \\
	&	&	& OFN & 0.615(14) & 0.306(2) & 0.356(3) & 0.415(8) & 0.463(11) & 0.585(6) \\
	&	&	&     &           & 0.366(6) & 0.394(7) & 0.387(5) & 0.412(5)  & 0.81(4)  \\
\hline
$7.3_I$ & 0.040 & 12	& OL  & 0.988(14)& 0.5481(7) & 0.799(13)& 0.861(6)& 0.883(9)& 0.735(2)\\
        &       &   	&     &          & 0.770(10) & 0.871(11)& 0.88(2) & 0.93(2) & 0.883(6)\\
	&	&	& OFN & 0.851(9) & 0.507(1)  & 0.575(3) & 0.618(3)& 0.652(4)& 0.734(4) \\
	&	&	&     &          & 0.586(4)  & 0.617(4) & 0.654(6)& 0.681(8)& 0.907(7) \\
\hline
$5.5_C$ & 0.030 & 12	& OL  & 0.845(9) & 0.4846(6) & 0.701(7) & 0.762(10) & 0.785(12) & 0.691(3) \\
        &       &   	&     &          & 0.697(7)  & 0.74(2)  & 0.741(8)  & 0.779(9)  & 0.851(8) \\
	&	&	& OFN & 0.725(10) & 0.430(2) & 0.486(3) & 0.546(7) & 0.580(6) & 0.668(4) \\
	&	&	&     &           & 0.484(4) & 0.519(4) & 0.525(3) & 0.543(6) &  0.93(4) \\
\hline
$7.5_I$ & 0.015 & 16	& OL  & 0.701(9) & 0.3387(7) & 0.500(8)  & 0.520(9)  & 0.551(4) & 0.502(4)\\
        &       &   	&     &          & 0.499(8)  & 0.526(6)  & 0.547(6)  & 0.57(2)  & 0.640(12)\\
	&	&	& OFN & 0.607(11)& 0.2892(12)& 0.337(2)  & 0.369(4)  & 0.389(4) & 0.484(3)\\
	&	&	&     &          & 0.336(3)  & 0.371(4)  & 0.389(5)  & 0.408(5) & 0.62(2) \\
\hline
$7.5_I$ & 0.030 & 16	& OL  & 0.826(15)& 0.4680(9) & 0.634(9)  & 0.676(4)  & 0.697(3) & 0.604(2)\\
        &       &   	&     &          & 0.621(4)  & 0.672(6)  & 0.687(7)  & 0.703(7) & 0.713(4)\\
	&	&	& OFN &  0.702(8)& 0.4076(9) & 0.4552(14)& 0.4824(16)& 0.505(2) &0.565(4) \\
	&	&	&     &          & 0.455(2)  & 0.481(3)  & 0.505(3)  & 0.524(3) &0.700(16)\\

\end{tabular}\end{center}
\newpage Table 1:
Summary of $\pi$ and $\rho_2$ masses from selected runs.
The top two lines of the table show the format for tabulating the masses.
The first column is the gauge
coupling, where the subscript ``C'' indicates the conventional
one-plaquette action and the subscript ``I'' indicates the Symanzik
improved gauge action.
The second column is the quark mass.
The third column is the
spatial lattice size, and the fourth column is the valence quark action.
In this
column ``O'' stands for the conventional ``one link'' action, ``OFN''
for the improved ``one-link + fat + Naik'' action.  ``OF'' and ``ON''
stand for the ``one-link + fat'' and ``one-link + Naik'' actions
respectively. The fifth column is the $\rho_2$ mass.
The remaining columns are pion masses.
The first four columns of pion masses correspond to distance 0, 1, 2 and 3 
operators respectively, all at zero momentum.  There are two
pion operators at each distance, shown one above the other.  These operators have the
``spin $\otimes
$ flavor'' structures shown in the table header \cite{GOLTERMAN_MESONS}.
The Goldstone pion is the upper pion in the distance zero column.
Goldstone pion energies at nonzero momenta are in the last column, labelled by the momentum in units
of $2\pi/N_s$.

\newpage
\begin{center}\begin{tabular}{llllll}
$\beta$ & $am_q$ & $S_V$ & $\frac{2\pi}{L m_\rho(0.55)}$ & $c^2(0,0,1)$ & $c^2(0,1,1)$ \\
\hline
7.1 & $0.1_4$ & OL & NA & 0.72(2) & 0.65(2) \\
7.1 & $0.1_4$ & OFN & NA & 0.98(4) & 0.99(2) \\
\hline

7.1 & 0.05 & OL &  0.53(2) & 0.82(1) & 0.75(1) \\	
7.1 & 0.05 & OFN & 0.57(1) & 0.94(2) & 0.95(3) \\	
\hline

7.3 & 0.02 & OL & 0.53(1) & 0.88(2) & 0.88(4) \\
7.3 & 0.02 & ON & --- & 0.97(2) & 0.93(3) \\
7.3 & 0.02 & OF & --- & 0.94(4) & 1.04(8) \\
7.3 & 0.02 & OFN & 0.64(1) & 0.96(5) & 1.09(10) \\
\hline

5.5 & 0.015 & OL & 0.66(1) & 0.95(5) & 0.91(4) \\
5.5 & 0.015 & ON & ---  & 0.98(5) & 0.97(4) \\
5.5 & 0.015 & OF & ---  & 0.88(6) & 1.10(9) \\
5.5 & 0.015 & OFN & 0.79(1) & 0.90(7) & 1.09(20) \\
\hline

7.3 & 0.04 & OL & 0.53(1) & 0.88(1) & 0.87(2) \\
7.3 & 0.04 & OFN & 0.64(1) & 1.03(3) & 1.03(2) \\
\hline

5.5 & 0.030 & OL &  0.66(1) & 0.88(2) & 0.89(3) \\
5.5 & 0.030 & OFN & 0.79(1) & 0.95(3) & 1.25(17) \\
\hline

7.5 & 0.015 & OL & 0.49(1) & 0.91(3) & 0.95(3) \\
7.5 & 0.015 & OFN & 0.60(1) & 0.93(5) & 0.95(3) \\
\hline

7.5 & 0.030 & OL &  0.49(1) & 0.92(1) & 0.94(3) \\
7.5 & 0.030 & OFN & 0.60(1) & 0.94(3) & 1.03(6) \\
\hline
\end{tabular}\end{center}
Table 2: ``Speed of light'' for the Goldstone pion.  The first three columns are the gauge coupling, quark
mass and valence quark action.  The fourth column is the unit of lattice momentum in units of the
arbitrary mass scale $m_\rho$ at $m_\pi/m_\rho=0.55$.  The last two columns are $c^2$ obtained
from the lowest two nonzero lattice momenta.
\newpage

\psfigure 5.0in 0.5in {OneLink} {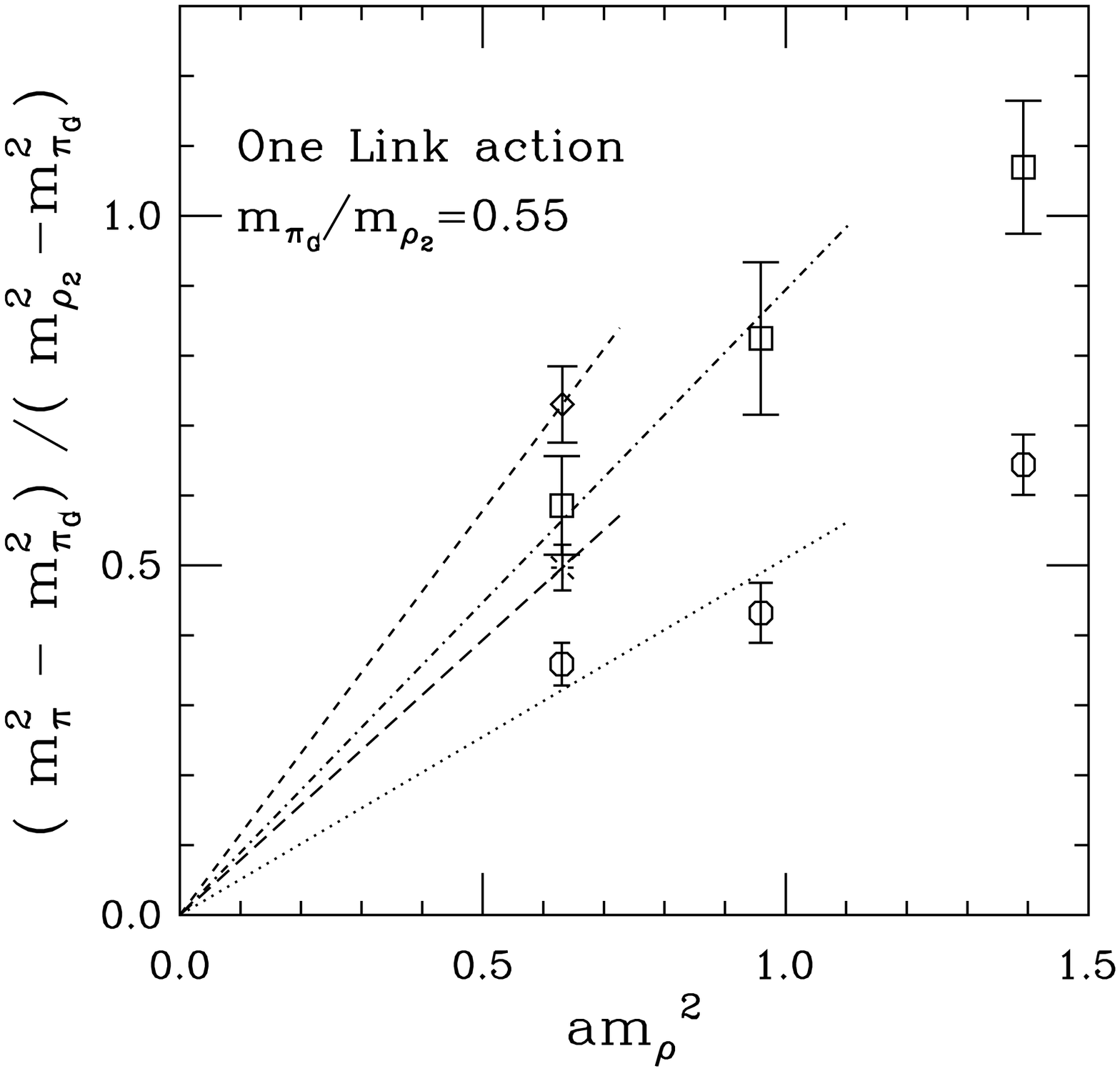} { $\delta_2$ for the
conventional (One Link) valence action.  Points with improved dynamical
action are at
$\beta_{imp}=7.1$, $7.3$ and $7.5$.  All the points are interpolated
to $m_\pi/m_\rho=0.55$ except for the $\beta_{imp}=7.1$ points for
which $m_\pi/m_\rho=0.51$. The octagons correspond to the local
non-Goldstone pion which has the smallest $\delta_2$ and the squares
correspond to the three-link $\gamma_0\gamma_5\otimes{\bf 1}$
pion which has the largest $\delta_2$.
Results with the conventional dynamical action at $\beta_{conv}=5.5$ are
also shown, where the burst is from the local non-Goldstone pion and the
diamond from the three-link pion.
The diagonal lines are crude extrapolations to smaller lattice spacing,
where the dotted line is the one+fat+Naik valence action local pion,
the dot-dashed line the one+fat+Naik valence action three link pion,
the long dashed line the conventional valence action local pion,
and the short dashed line the conventional valence action three link pion.
}

\psfigure 5.0in 0.5in {OneFatNaik} {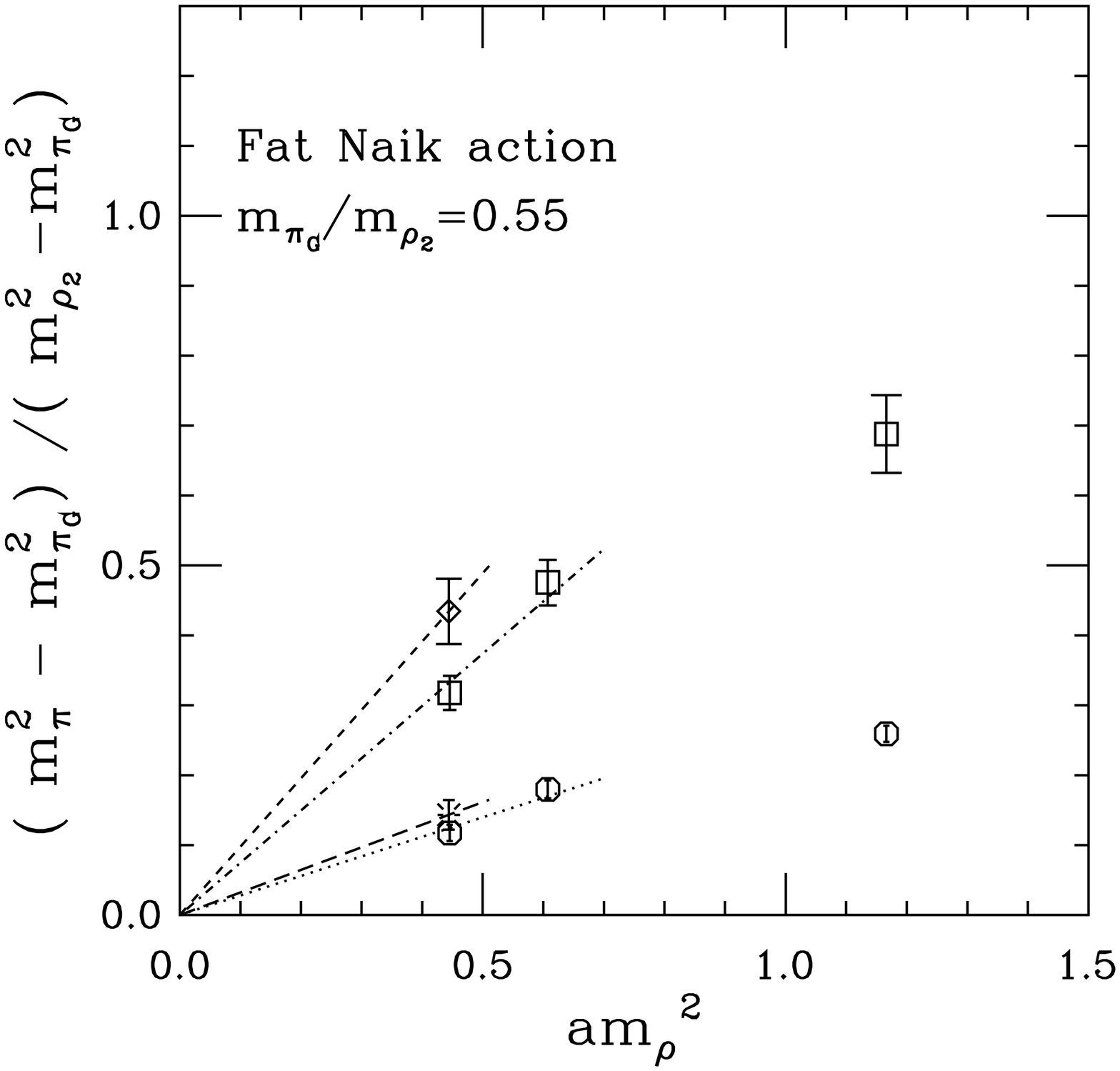} { $\delta_2$
for the fat+Naik valence action.
The dynamical actions and the symbols used in plotting
are the same as in Fig.~\protect\ref{OneLink}.
All the points are interpolated
to $m_\pi/m_\rho=0.55$ except from the the $\beta_{imp}=7.1$ points for
which $m_\pi/m_\rho=0.57$.
}
Another symmetry which is broken on the lattice is the flavor
symmetry.  The pion masses with $\beta_{imp}=7.3$ or
$\beta_{conv}=5.5$ with various valence quark actions in Table~I show
that the fattening of the link improves the flavor symmetry while the
Naik term has little effect, in agreement with earlier studies in the
quenched approximation\cite{MILC_NAIK,MILC_FATLINKS,ILLINOIS_FAT}.
Here we also see that the improvement extends to the one-link nonlocal
pions.  We can parameterize the splitting of the pions by the
dimensionless quantity 
\BE 
\delta_2 = \frac{M_\pi^2 -M_G^2}{M_\rho^2-M_G^2} \ \ \ .
\EE 
Here $M_\pi$ is one of the
non-Goldstone pion masses, $M_G$ is the Goldstone pion mass, and
$M_\rho$ is one of the local $\rho$ masses.  Since the $\rho$ masses are
nearly degenerate, it makes little difference which one we use.  
In our analysis we used the $\rho_2$ mass because it is usually
estimated more accurately.  
This measure of flavor symmetry breaking differs from the quantity used in
\cite{MILC_FATLINKS}.  Here we use the squared masses since they are
approximately linear in the quark mass and we will do some
interpolating and extrapolating.  Also, empirically this $\delta_2$ is
fairly insensitive to the quark mass.  The squared Goldstone pion mass is
subtracted in the denominator to give more sensible behavior at large
quark mass.  At large $am_q$ all of the meson masses are becoming
degenerate, but it is still possible to use the difference between the
Goldstone pion mass and the rho mass as a natural scale for flavor
symmetry breaking.  The local non-Goldstone pion is the easiest to
measure, and one for which results with the conventional action are
available in the literature.  Figures~\ref{OneLink} and \ref{OneFatNaik}
show $\delta_2$ for some of the pions as a function of lattice spacing,
both for this improved action (both the dynamical and valence quark
actions are improved), and for the conventional
action.
We show $\delta_2$ using the local non-Goldstone pion
($\gamma_0\gamma_5\otimes\gamma_0\gamma_5$), which is consistently the
lightest non-Goldstone pion, and the distance three
$\gamma_0\gamma_5\otimes{\bf 1}$ pion, which is usually the heaviest.
Results for the other pion operators can be filled in using the values
in Table I.
Here we have plotted results for both conventional and improved dynamical
actions in the same plot, with the valence actions in separate plots.
Thus, to compare a completely conventional simulation (both valence and dynamical
actions) with a simulation with both valence and dynamical actions improved,
one should compare the burst and diamond in Fig.~\ref{OneLink} with the octagons and
squares in Fig.~\ref{OneFatNaik}.
In these figures all of the
data have been interpolated to $m_\pi/m_\rho=0.55$ except for the
points at $\beta=7.1$.  

Clearly, the fat+Naik action shows a significant improvement of the
flavor symmetry breaking over the conventional action.  This
improvement is due to the ``fattening'' and not due to the Naik term
which only affects the dispersion relation.  The nonlocal pions have
significantly larger splittings than those of the local
$\pi_2$. Although their splittings get reduced along with the $\pi_2$
splitting, it is evident that the reduction is smaller.  However, as
expected, all the pions are approaching the same mass as the lattice
spacing gets small. From our results, it is clear that one has to
consider the nonlocal pions before making any judgment about the
quality of the improved action.

Furthermore, it is evident that part of the flavor symmetry
restoration is due to the improvement of the gauge action. This is
also supported by the quenched spectroscopy results presented
in\cite{MILC_FATLINKS}.  In the case of the one link action
(Fig.~\ref{OneLink}), the improvement of the gauge action
significantly reduces both the local and the nonlocal pion
splittings. The local pion splittings are affected more than those of
the nonlocal.  In the case of the fat+Naik action
(Fig.~\ref{OneFatNaik}), we also see a reduction of all the
splittings. However, the local pion splittings are improved very
little while the nonlocal get a major improvement.  From the above
observations one may conclude that the ``fattening'' of the fermion
action is less effective for the the nonlocal pions which they get
improved mostly because of the the gauge action improvement.  Further
``fattening'' may be needed in order to significantly improve the
nonlocal pions.

\section{Conclusions}

%
%
%
%
%
%

The first conclusion is that, as expected,
the improvements in rotational symmetry and
flavor symmetry seen in the quenched spectrum also appear in full QCD
spectrum calculations.
The most serious problem for accurate simulations of full QCD with
Kogut-Susskind fermions is the flavor symmetry breaking.  While breaking
of rotational symmetry is also serious at large lattice spacings, the
light meson dispersion relations
are in much better shape than the flavor symmetry.  In particular, in our
simulations with $\beta_{imp}=7.5$ the speed of light for the pions is quite
acceptable, while the flavor symmetry is still seriously broken.

We have extended the earlier studies of flavor symmetry breaking to include
the nonlocal pions as well as the local non-Goldstone pion.  The
splittings between the nonlocal pions and the Goldstone pion are also reduced
by improving the action, but they are significantly more massive than
the local non-Goldstone pion.
Worse, the amount of flavor symmetry improvement from the fat link
used here is less for the nonlocal pions than the local one.
This is seen in Figs.~\ref{OneLink} and \ref{OneFatNaik}, where we have drawn some lines
crudely indicating $\delta_2$ for the local pion and the 
3-link $\gamma_0\gamma_5\otimes{\bf 1}$ pion, which is consistently
the heaviest pion.
Note that the slope of the line for the local pion was
reduced by a larger fraction than was the slope for the 3-link pion when
the valence action was improved, {\it i.e.} in going from Fig.~\ref{OneLink} to
Fig.~\ref{OneFatNaik}.  Unfortunately, in a realistic simulation we would want all of
the pions to be well behaved, so in estimating the lattice spacing
required for the desired quality simulation we must consider the
nonlocal pions.

An important use of action tests such as this work is to estimate
the lattice parameters required to investigate real problems in QCD.
An important example, and the example which partially motivated this
work, is the study of the high temperature QCD phase transition.
Interesting questions here include an accurate determination of the
equation of state\cite{MILC_EOS},
and the question of whether the addition of a strange
quark changes the nature of the crossover or phase
transition\cite{ADD_STRANGE}.
To be sure that we are doing an accurate simulation in the crossover
temperature range, we would like to know that our simulation is under
control on both the high and low temperature sides of the crossover.
The addition of the Naik term does an excellent job of improving the
quark dispersion relation, and hence we expect to improve
the equation of state in the high
temperature phase\cite{BIELEFELD_NAIK}.  However, the harder part of the
job is to get the physics right on the low temperature side.  Roughly,
there the physics is a gas of thermally excited pions.  To reproduce
this physics, we need the pion dispersion relation to be good up to
energies large compared to the temperature, and the pion masses to be
nearly degenerate --- that is, for flavor symmetry to be restored.
For concreteness, suppose that we are interested in the effects of
adding a strange quark.  Viewed from the low temperature side of the
transition, this adds kaons with mass 500 MeV to the physics.  To
isolate this effect, we should require that all of the pions have
masses small compared to this.  For example, we might demand that when
the Goldstone pion has its physical mass of 140 MeV, that the
non-Goldstone pions have masses less than one half the kaon mass, or 250
MeV.  This requires $\delta_2 \le 0.075$.
The $\delta_2$'s for the various pions are
fairly insensitive to quark mass. ( In the $10/g_I^2=7.5$
runs they are equal within errors at $am_q=0.015$ and $am_q=0.030$ for both
the one-link and one-fat-Naik valence actions.)
Neglecting dependence of $\delta_2$ on quark mass, it is straightforward to
find the necessary value of $a m_\rho$ from the lines in Figs.~\ref{OneLink}
and \ref{OneFatNaik}, and from there to figure out
the number of time slices needed to
simulate a temperature of 150 MeV.  Requiring that the heaviest pion
mass be less than 250 MeV when the Goldstone pion is at 140 MeV
will require about 20 time slices with the conventional action, and
about 16 with the one+fat+Naik action.  (Had we used the local pion
in this exercise, the results would have been 16 and 10 time slices
respectively.)  Sixteen time slices is a disappointingly small lattice
spacing, so it is important to extend these tests to actions which might
further reduce flavor symmetry breaking, such as the action discussed in
Ref.~\cite{ILLINOIS_FAT}.

\section*{Acknowledgements}
This work was supported by the U.S. Department of Energy under contract
DE-FG03-95ER-40906. 
Computations were done on the CPPACS pilot machine at the
University of Tsukuba, the Paragon at Oak Ridge National
Laboratory and the T3E at NERSC.
D.T. is grateful for the hospitality of the 
University of Tsukuba, where this work was begun.
We thank Tom DeGrand and Anna Hasenfratz for their
generalized gauge force code for the molecular dynamics algorithm.
We would like to thank Graham Boyd, Tom DeGrand and Urs Heller for helpful
discussions.


\end{document}